\documentstyle[preprint,eqsecnum,aps]{revtex}

\begin{document}

\count255=\time\divide\count255 by 60 \xdef\hourmin{\number\count255}
  \multiply\count255 by-60\advance\count255 by\time
  \xdef\hourmin{\hourmin:\ifnum\count255<10 0\fi\the\count255}

\preprint{\vbox{\hbox{WM-98-119}\hbox{JLAB-THY-98-50}
}}

\title{Operator Analysis of $\ell = 1$ Baryon Masses in Large $N_c$ QCD}

\author{Carl E. Carlson$^\dagger$, Christopher D.  Carone$^{\dagger}$,
Jos\'{e} L. Goity$^{\ddagger\diamond}$, and Richard
F. Lebed$^\diamond$}

\vskip 0.1in

\address{$^\dagger$Nuclear and Particle Theory Group, Department of
Physics, College of William and Mary, Williamsburg, VA 23187-8795\\
$^\ddagger$Department of Physics, Hampton University, Hampton, VA
23668 \\ $^\diamond$Jefferson Lab, 12000 Jefferson Avenue, Newport
News, VA 23606}

\vskip .1in
\date{December 1998}
\vskip .1in

\maketitle

\begin{abstract}
We consider in detail the mass operator analysis for the nonstrange 
$\ell=1$ excited baryons in large $N_c$ QCD.   We present a straightforward
procedure for constructing the large $N_c$ baryon wavefunctions, and 
provide complete analytic expressions for the matrix elements of all 
the independent isosinglet mass operators.  We discuss the relationship
between the old-fashioned operator analyses based on nonrelativistic
SU(6) symmetry and the modern large $N_c$ approach, which has a firmer 
theoretical foundation.  We then suggest a possible dynamical 
interpretation for the subset of operators preferred strongly 
by the data.
\end{abstract}

\thispagestyle{empty}

\newpage
\setcounter{page}{1}

\section{Introduction}

Although the QCD gauge coupling is numerically too large to permit a
perturbative expansion at low energies, QCD generalized to $N_c$ colors 
admits a consistent perturbative expansion in terms of $1/N_c$\cite{tHooft}.  
Effective theories for baryons have been constructed that take into 
account the symmetries and power counting rules of large $N_c$ QCD, allowing 
baryon observables to be computed to any desired order in the $1/N_c$ 
expansion.  The large $N_c$ approach has been applied with great success 
to the ground state baryons which fill the SU(6) {\bf 56}-plet, including 
studies of SU(6) spin-flavor symmetry \cite{DM,Jenk1,DJM1,CGO,Luty1}, baryon 
masses \cite{DJM1,Jenk2,DJM2,JL}, magnetic 
moments~\cite{DJM1,DJM2,JL,JM,Luty2,DDJM}, and axial current matrix
elements \cite{DM,DJM1,DJM2,DDJM}.

Whether the large $N_c$ framework works equally well in describing the
phenomenology of excited baryon multiplets is a question under active
investigation.  Recent attention has focused on the $\ell=1$
orbitally-excited baryons, the SU(6) {\bf 70}-plet for $N_c=3$.  The
first application of large $N_c$ to excited baryons was a
phenomenological analysis of the strong decays \cite{CGKM}.  This was
followed by a series of more formal papers on the strong decays and
axial current matrix elements \cite{PY2,PY1}, as well as on the matrix
elements of the mass operators relevant at lowest nontrivial
order~\cite{Goity}.  Recently, the first phenomenological study of the
electromagnetic transitions was presented \cite{CC,CC2}, while a
phenomenological analysis of the nonstrange $\ell=1$ baryon masses,
including corrections up to relative order $1/N_c^2$, was undertaken
by the present authors \cite{CCGL}.  This is the subject of further
consideration in the the present work.

A number of issues not addressed in Ref.~\cite{CCGL} are considered
here.  First, we explain how the nonstrange baryon states are
constructed for arbitrary $N_c$.  Our construction differs from that
of Ref.~\cite{PY2}, and we believe is somewhat more transparent.
After obtaining rules for simplifying the baryon operator analysis,
which is essential for a proper counting of degrees of freedom in fits
to observables, we present complete analytic expressions for the
matrix elements of all isosinglet mass operators
relevant to the orbitally-excited baryons, as functions of the excited
baryon quantum numbers.  This presentation is relevant not only for
obtaining the large $N_c$ results presented in our earlier work but
also for identifying operator relationships holding only to leading
order in $1/N_c$.  For example, the matrix elements of some of the
operators are linearly dependent in the $N_c\rightarrow \infty$ limit,
even though the matrix elements are independent for $N_c=3$.  Thus,
two operators that appear naively to be of leading order in $1/N_c$
may in fact produce only one leading-order linear combination.  The
operator basis presented here is thus slightly improved over that of
Ref.~\cite{CCGL}.  We present numerical results omitted from 
Ref.~\cite{CCGL} for reasons of space, namely, 
fits to mass eigenvalues in which the mixing angles are
predicted.  We also consider the physical interpretation of our
effective field theory results.  It was shown in Ref.~\cite{CCGL} that
only two nontrivial operators have numerically substantial
coefficients when fits to the nonstrange $\ell=1$ mass spectrum are
performed, and this in itself is suggestive of some specific dynamical
mechanism.  In this work we attempt to characterize the dynamics 
producing these results.

This paper is organized as follows.  In Sec.~2 we review the
formulation of the large $N_c$ operator analysis for excited baryons.
In Sec.~3 we describe in detail the construction of the baryon states
in large $N_c$.  Sections~4 and 5 discuss operator reduction rules and
construction of the operator basis relevant to the mixed-symmetry 
{\bf 70}-plet states.  In Sec.~6 we present numerical results not 
included in our prior work.  In Sec.~7 we compare our results to 
model-independent analyses of the past, and in Sec.~8 to phenomenological 
models.  Section~9 summarizes our conclusions.

\section{Framework}

The observed baryons have the appropriate quantum numbers to be
assigned to irreducible representations of the group
SU(6)$\times$O(3).  Here SU(6) contains the spin and flavor symmetry
group SU(2)$\times$SU(3), and O(3) generates spatial rotations.  We
define ``quarks'' $q$ as fields in the ({\bf 2},{\bf 3})
representation of the spin-flavor group.  An appropriately symmetrized
collection of $N_c$ quarks has the quantum numbers of a large $N_c$
baryon.  For $N_c=3$, states constructed in this way have the
same quantum numbers as those observed in nature.

If all quarks were much heavier than $\Lambda_{{\rm QCD}}$, then one could
identify the fields above as the valence quarks of the nonrelativistic
quark model.  Here, however, we make no such assumption.  Our quark
fields simply provide a convenient tensor product space in which one
can define baryons with the correct total quantum numbers.  The baryon
wavefunctions can be expressed as tensors, with separate indices for
the spin and flavor degrees of freedom for each quark.  In the
nonrelativistic quark model, all spin-flavor transformations of the
baryon tensors are accomplished by acting on these indices with elements
of the group SU(6), which is an exact symmetry of the theory in the
limit $m_q \rightarrow \infty$, where $m_q$ is the quark mass.  In the
present case, we cannot (and do not) assume that SU(6) is a good
symmetry, since the quarks are light, but rather simply parameterize
the complete breaking of SU(6) by allowing symmetry-breaking matrices
to act on the quark spin and flavor indices.  One achieves the most
general breaking of quark spin and flavor symmetries by using polynomials 
in the SU(6) generators
\begin{equation}
\left( \frac{\sigma^i}{2} \otimes \openone \right), \ \
\left( \openone \otimes \frac{\tau^a}{2} \right), \ \
\left( \frac{\sigma^i}{2} \otimes \frac{\tau^a}{2} \right),
\label{eq:su6gen}
\end{equation}
where $\sigma^i$ are the usual Pauli matrices. The $\tau^a$ are either
Pauli or Gell-Mann matrices, depending on whether one is interested in
two or three quark flavors.  We focus on the two-flavor case in our
operator analysis.  By acting on the quark spin and flavor indices of a
baryon wavefunction, the tensors above parameterize the breaking
of the corresponding symmetries. Within a large $N_c$ baryon multiplet, 
there are {\em  always} some states for which these symmetry breaking 
effects are maximal. For example, consider the ground state baryons, which 
form a tower of states with spins ranging from $1/2$ to $N_c/2$.  The fact
that the large $N_c$ multiplet contains states with spins of order
$N_c$ implies that spin-spin interactions like
\begin{equation}
  \frac{1}{N_c} S^2 \equiv \frac{1}{N_c} \sum_{\stackrel{\rm
      quarks}{\alpha,\beta}} \frac{\vec{\sigma}_\alpha}{2} \cdot 
\frac{\vec{\sigma}_\beta}{2}
\end{equation}
shift some baryon mass eigenvalues at order $N_c$. (The reason for the
$1/N_c$ prefactor is explained below.) For example, for
the stretched case of a baryon with spin $N_c/2$, this matrix element
evaluates to $1/N_c \cdot N_c/2 \cdot (N_c/2 + 1)$.  On the other hand, 
the mean mass of the multiplet scales as $N_c$, since there are $N_c$ quarks in
a baryon state.  Thus there are always
spin-dependent splittings somewhere in the multiplet that are
comparable to the average multiplet mass.  While this prevents us from
speaking of SU(6) as an approximate symmetry, it is nonetheless true
that the breaking of this would-be symmetry is a small effect on
states of small total spin.  Since the physical, nonstrange
baryons are chosen to have fixed total spin and isospin eigenvalues in
the large $N_c$ limit, it follows that matrix elements of $\sigma^i/N_c$
and $\tau^a/N_c$ summed over all quarks are of order $1/N_c$, and hence 
can be treated as small numbers.  Thus, this parameterization of the complete 
breaking of SU(6) provides an operator basis that is hierarchical in $1/N_c$ 
on the physical baryon states.  This fact allows the construction of an 
effective theory for baryons that is both complete and predictive.

Let us now specify the large $N_c$ counting rules more precisely.  We
define an $n$-body operator as one that acts on $n$ quark lines in a
large $N_c$ baryon state.  Since we work in an effective theory, we arrive 
at a complete operator basis independent of any specific dynamical 
assumptions beyond that of QCD as the underlying theory.  An $n$-body
operator has a coefficient $1/N_c^{n-1}$, reflecting the minimum $n-1$ gluon
exchanges necessary to generate the operator in QCD.  However, the
overall effect of an operator on a given baryon observable is
determined not only by the size of the operator coefficient, but also
by the compensating factors of $N_c$ that may arise when a spin-flavor
generator is summed over the $N_c$ quark lines in a baryon state.  As
discussed earlier, the generators $\sigma^i$ and $\tau^a$ sum incoherently
over $N_c$ quark lines since the spin and isospin eigenvalues for the
physical baryons are of order one, even when one extrapolates to large
$N_c$.  The generator $\sigma^i\tau^a$, however, sums coherently, as is 
shown later by explicit computation [See Eq.~(\ref{eq:giame})].  Thus, 
the contribution of an $n$-body operator to a given baryon observable 
is of order $N_c^{1+m-n}$, where $m$ is the number of times the generator 
$\sigma^i\tau^a$ appears.  Given the set of all operators constructed by 
combining the generators in (\ref{eq:su6gen}), linearly dependent operators 
of higher order can often be eliminated by use of operator reduction rules.  
For the ground state baryons, these rules were formalized by Dashen, Jenkins
and Manohar\cite{DJM2}; the generalization to excited baryons is
considered in some detail in Section~IV.

The discussion above generalizes in a straightforward way to $\ell=1$
baryons with one orbitally-excited quark.  In the large $N_c$ limit,
such baryons consist of one distinguishable, excited quark in the
collective potential generated by $N_c-1$ ground state quarks.  One
defines separate SU(6) generators that act on the excited quark, and
on the non-excited  ``core'' quarks, respectively.    In addition, one
introduces the orbital angular momentum generators $\ell^i$ to
parameterize the breaking of O(3).  Mass operators relevant to the 
$\ell=1$ baryons are formed by contracting generators in this extended set,
as we discuss in Section~IV.  Again, an operator hierarchy is
obtained after taking into account the factors of $1/N_c$ that appear
in operator coefficients, and the compensating factors of $N_c$ that arise
from coherent sums over the $O(N_c)$ ground state core quarks.

\section{States}

The defining feature of baryon states filling the mixed-symmetry
negative-parity SU(6) {\bf 70}-plet is that the sole unit of
orbital angular momentum is carried by the excited quark relative to the 
other two ground-state core quarks. The core quarks are separately 
symmetrized on spin-flavor and spatial indices, while 
the $\ell=1$ excited quark is antisymmetrized with respect to the other two.
This construction produces the {\bf 70}-dimensional representation 
of SU(6), and is phenomenologically relevant: Every negative-parity baryon 
with mass less than 2 GeV has the appropriate spin, isospin, and strangeness 
quantum numbers to belong to a single {\bf 70}-plet, although some of 
the strange baryons needed to fill the {\bf 70} have not yet been observed.
If one focuses upon nonstrange states alone, as is done in this work, then 
the relevant multiplet becomes a {\bf 20} of SU(4), for which all component 
spin and isospin multiplets have been seen.

\def\nboxF{\vbox{\hbox{$\bsqr\bsqr\bsqr\bsqr\raise5.0pt\hbox{$\,\cdot
\cdot\cdot\cdot\cdot\,$}\bsqr\bsqr$}\nointerlineskip 
\kern-.2pt\hbox{$\bsqr$}}}

\begin{figure}[h]
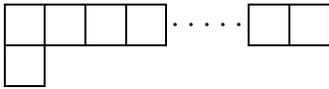

  \begin{raggedright}

\def\ssqr#1#2{{\vbox{\hrule height #2pt
      \hbox{\vrule width #2pt height#1pt \kern#1pt\vrule width #2pt}
      \hrule height #2pt}\kern- #2pt}}
\def\sqr{\mathchoice\ssqr8{.4}\ssqr8{.4}\ssqr{5}{.3}\ssqr{4}{.3}}

\def\bsqr{\ssqr{15}{.3}}

\centerline{$$\nboxF$$}
\bigskip

\caption{Young diagram for the SU($2F$) mixed symmetry representation,
  the multiplet containing large $N_c$ orbitally-excited baryons with
  $\ell = 1$.  The top row has $N_c - 1$ boxes.}
\label{young}

  \end{raggedright}
\end{figure}

The mixed symmetry baryon multiplet is generalized to $N_c >3$ by
symmetrizing now among $N_c-1$ core quarks, as indicated by the Young
diagram in Fig.~1.  Although this extrapolation is not unique, it is
the most natural in preserving symmetry properties familiar from
$N_c=3$.  Total symmetry of the core is also the essential ingredient
rendering the study of the orbitally excited baryons tractable in
large $N_c$, since it greatly reduces the number of degrees of
freedom.  In particular, the symmetry properties of core states are
completely specified by their total strangeness, spin, and isospin.
For nonstrange cores, the situation is even simpler: Owing to the
total symmetry of the spin-flavor state, spin and isospin are equal in
this case.  The core state is denoted by
\begin{equation}
\left|S_c=I_c;m_1,\alpha_1 \right> ,
\end{equation}
where $m$'s and $\alpha$'s here and below denote projections of spin
and isospin, respectively, and the subscript $c$ denotes core.  The
excited quark state is denoted
\begin{equation}
\left|1/2;m_2,\alpha_2 \right> .
\end{equation}
Finally, the orbital O(3) eigenstate is labeled in obvious notation by
\begin{equation}
\left|  \ell,m_\ell \right>.
\end{equation}

Of course, physical states are labeled by total spin $J, J_3$ and
isospin $I, I_3$.  The states we construct here also admit separate 
specification of the total spin $S$ carried by the quarks.  Nonstrange
mixed-symmetry SU(6) states with one spin-1/2, isospin-1/2 quark singled 
out have total quark spin and isospin related by $S=I$ or $I \pm 1$, with each
of $S$ and $I$ in the range $1/2$ to $N_c/2$.  The sole exception is
that there are no mixed-symmetry $S=I=N_c/2$ states.  Let us 
define $\rho \equiv S-I = \pm 1, 0$, and $\eta/2 \equiv I_c - I = \pm 1/2$.  
Then obtaining the desired state by coupling the spins and isospins is
achieved, by construction, by the use of Clebsch-Gordan coefficients:
\begin{eqnarray}
\left| J J_3; I I_3 \, (\ell, \, S=I+\rho) \right\rangle &=&
  \sum_{m_\ell,m_1,\alpha_1,\eta}
                  \left(
                        \begin{array}{cc|c}
                             \ell & S   & J \\
                             m_\ell & m & J_3
                        \end{array}
                                          \right)
                  \left(
                        \begin{array}{cc|c}
                             S_c & 1/2   & S \\
                             m_1 & m_2   & m
                        \end{array}
                                          \right)
                  \left(
                        \begin{array}{cc|c}
                          I_c      & 1/2      & I \\
                          \alpha_1 & \alpha_2 & I_3
                        \end{array}
                                          \right)
      c_{\rho,\eta}
                         \nonumber \\[2ex]
&\times&
          \left|S_c=I_c=I+\eta/2;m_1,\alpha_1 \right>
\otimes
          \left|1/2;m_2,\alpha_2 \right>
\otimes
          \left|  \ell,m_\ell \right> \label{states}
\end{eqnarray}
States with strangeness are defined analogously, except that SU(3)
Clebsch-Gordan coefficients appear in that case.  The only notation in
this expression yet undefined is the coefficient $c_{\rho \eta}$; it
simply represents that more than one irreducible SU(6) or SU(4)
representation can occur in the product of the ($N_c-1$)-quark core and
the one-quark excited state, and so the numbers $c_{\rho \eta}$
represent elements of orthogonal basis rotations.  In the present
case, elementary manipulations show that only the totally symmetric
and mixed-symmetry representations result.  Since $S-S_c = \pm 1/2$, 
one has $c_{\pm, \mp} = 0$ for any multiplet.  All nonstrange states in the
symmetric representation have $S=I$, and thus $c^{\rm SYM}_{\pm, \pm} = 0$, 
$c^{\rm MS}_{\pm,  \pm} = 1$.  The only complicated mixing occurs 
for $c_{0, \pm}$, and we obtain the mixing by means of a trick: The symmetric 
and mixed-symmetry multiplets possess different quadratic SU(4) or SU(6) 
Casimirs, and thus one may compute the value of the Casimir both on the 
full state on the left-hand side of (\ref{states}), where it assumes a 
known value (see next section), or on the separate core and excited states 
on the right-hand side of (\ref{states}) using the matrix elements presented 
in Appendix~\ref{sec:apa}.  After a straightforward calculation, one finds 
for the $S=I$ nonstrange states
\begin{eqnarray}
c^{\rm MS}_{0+} = + \sqrt{S \, (N_c+2(S+1)) \over N_c \, (2S+1) }
  \quad {\rm and} \quad
c^{\rm MS}_{0-} = - \sqrt{(S+1)(N_c-2S) \over N_c \, (2S+1) },
\end{eqnarray}
and the coefficients for symmetric states are the orthogonal combination,
$c^{\rm SYM}_{0+} = -c^{\rm MS}_{0-}$, $c^{\rm SYM}_{0-} = c^{\rm
  MS}_{0+}$.

\section{Reductions}

There are numerous operator identities or operator reduction
rules which are known for the ground state baryons and which
can be used to eliminate many operator products from lists of
candidate independent operators.  The identities are not
general to all representations, but work when applied to ground
state baryons.  The proofs of many of them depend upon the
symmetry of the ground state.

In this section we study operator reductions applicable to the
mixed-symmetry {\bf 70}-plet.  Technical details are provided
in Appendix~\ref{sec:apb}. To put our findings in context,
recall that the operator reductions for the ground state come
from three sources.  Two of them are the quadratic and cubic
Casimir identities.  The third comes because matrix elements of
an operator between a state and its conjugate state are zero if
the operator does not belong to a representation that can be
found in (for the ground state with $N_c=3$)
$\overline{{\bf 56}} \otimes {\bf 56}$.  There are products of two
generators of SU(6), or rather certain sums of products of
these generators, that belong to representations not found in
$\overline{{\bf 56}} \otimes {\bf 56}$.  Those sums are then zero, and this
is the third source of operator identities.  Ref.~\cite{DJM2}
investigates whether further identities can be found involving
products of three  generators, and shows that the answer is
negative.  

The basic operators that we start with are the
generators of SU($2F$), given in a quark basis as
\begin{eqnarray} \label{basis}
S^i & \equiv & q^\dagger_\alpha 
  \left( \frac{\sigma^i}{2} \otimes \openone \right) 
    q^\alpha , \nonumber \\  
T^a & \equiv & q^\dagger_\alpha 
  \left( \openone \otimes \frac{\tau^a}{2} \right)
    q^\alpha , \nonumber \\ 
G^{ia} & \equiv & q^\dagger_\alpha 
  \left( \frac{\sigma^i}{2} \otimes \frac{\tau^a}{2} \right)
    q^\alpha ,
\end{eqnarray}
where $\sigma^i$ and $\tau^a$ are the spin and flavor matrices. 
The collected and properly normalized SU($2F$) adjoint
representation one-body operators $S^i/\sqrt{F}$,
$T^a/\sqrt{2}$, and $\sqrt{2} G^{ia}$ satisfy an SU($2F$)
algebra like that of their underlying spin-flavor
generators.\footnote{The normalizations are chosen so that the
underlying spin-flavor generators $\Lambda^A \equiv 
\left\{ \left( \sigma^i/2 \otimes \openone \right)
                                      /\sqrt{F}, \, 
\left( \openone \otimes \lambda^a/2 \right) /\sqrt{2}, \, 
             \sqrt{2} 
\left( \sigma^i/2 \otimes \lambda^a/2 \right) \right\}$ 
satisfy 
${\rm Tr} \, \Lambda^A \Lambda^B = \frac{1}{2} \delta^{AB}$.}
Other operators $\cal O$ can, since
we are just interested in their group theoretical behavior, be
built from products of these generators~\cite{DJM2}.

For the {\bf 70}-plet, first note that the mixed-symmetry
representation consists of a symmetric core plus one excited
quark.   If one defines, in analogy with (\ref{basis}),
separate one-body operators $S_c$, $T_c$, $G_c$ acting on the
core and $s$, $t$, $g$ on the excited quark line, then the
operator reduction rules for the ground state \cite{DJM2} may
be used on the core operators. The only difference is that $N_c
\to N_c-1$ in the core identities, to account for the different
numbers of quarks present.

For the {\bf 70}-plet overall, we find that the quadratic
Casimir leads to a new operator reduction rule. 
Unfortunately, the other two sources of identities for the
ground state lead to no identities for the {\bf 70}-plet. 
However, there are some identities that come from considering
products of three currents.

The quadratic Casimir identity for an arbitrary SU($2F$)
representation $R$ reads
\begin{equation} \label{quad}
\left\{ q^\dagger \Lambda^A q, \, q^\dagger \Lambda^A q \right\}
\equiv 2C_2(R) \, \openone ,
\end{equation}
where $\Lambda^A$ are the spin-flavor generators in the
representation $R$.  For the mixed-symmetry representations we
are looking at, denoted MS$_{N_c}$, the Casimir may be
shown to be (see~\cite{gt93})
\begin{equation} C_2 \left( {\rm MS}_{N_c} \right) =
\frac{N_c}{4F} \left[ N_c (2F-1) + 2F(2F-3) \right] .
\end{equation}
In the mixed-symmetry case, $\Lambda^A$ is the sum of core and
excited generators ({\it i.e.}, $T = T_c + t$, {\it etc.}), and
\begin{eqnarray} 
C_2 ({\rm S}_1) & = & \frac{1}{4F} (4F^2-1) , 
\nonumber \\ 
C_2 ({\rm S}_{N_c-1}) & = & 
\frac{1}{4F} (N_c-1) (N_c+2F-1) (2F-1) .
\end{eqnarray}
where S$_{N_c-1}$ is the symmetric representation with $N_c-1$
quarks and S$_1$ is just the fundamental representation
of a single quark.  This means that the quadratic
Casimir identity for MS$_{N_c}$ can be expressed as,
\begin{equation} 
2 s S_c + F t T_c + 4 F g G_c 
     = -\frac{1}{2} (N_c + 2F - 1) ,
\end{equation} 
so that the operator $gG_c$ may always be eliminated in
favor of $sS_c$ and $tT_c$.

The cubic Casimir identity reads,
\begin{equation} \label{truecubic}
d^{ABC} 
    \left( q^\dagger \Lambda^A q \right) 
    \left( q^\dagger \Lambda^B q \right) 
    \left( q^\dagger \Lambda^C q \right) 
\equiv C_3 (R) \, \openone .
\end{equation}
For products of two generators contracted with $d^{ABC}$, one
may write
\begin{equation} \label{cubic}
d^{ABC} 
    \left( q^\dagger \Lambda^B q \right) 
    \left( q^\dagger \Lambda^C q \right)
\equiv 
    \frac{C_3 (R)}{C_2(R)} \, 
           q^\dagger \Lambda^A q 
                                  + X^A(R) ,
\end{equation}
where $X^A$ is that part of the two-body combination
on the left hand side annihilated by contraction with 
$q^\dagger \Lambda^A q$.  For completely
symmetric representations $X^A = 0$, as was shown explicitly in
\cite{DJM2};  one can show the same for completely
antisymmetric representations. In such cases, one may
derive a number of operator reduction rules. However, $X^A$ need
not be zero for arbitrary representations, since nothing
guarantees that all spin-flavor combinations of the quark
operators
$(q^\dagger q) (q^\dagger q)$ reduce to a single $(q^\dagger
q)$ for a representation of arbitrary symmetry properties.  We
have found explicitly that $X^A \neq 0$ for the mixed-symmetry
representation by computing several matrix elements containing
both sides of (\ref{cubic}).  One concludes that no
two-body operator reduction rules follow from the cubic
Casimir relation for the mixed-symmetry representation.

Of course, the true cubic Casimir relation~(\ref{truecubic})
holds in general.  We have investigated it for the
mixed-symmetry representation, and find no new operator
relations, but rather the Casimir identity
\begin{equation}
              \frac{2}{3} \left[ 
\frac
 {C_3 ({\rm MS}_{N_c}) - 
             C_3 ({\rm S}_{N_c-1}) - C_3({\rm S}_1)} 
 {C_2 ({\rm MS}_{N_c}) - 
             C_2 ({\rm S}_{N_c-1}) - C_2({\rm S}_1)} 
                                 \right] 
= 
    \frac{C_3({\rm S}_{N_c-1})}{C_2({\rm S}_{N_c-1})} 
+ 
    \frac{C_3 ({\rm S}_1)}{C_2({\rm S}_1)} ,
\end{equation}
which can indeed be verified, using the previous quadratic
Casimirs and
\begin{eqnarray}
C_3 ({\rm MS}_{N_c}) & = & \frac{N_c}{4F^2} (F-1) (N_c +2F)
              \left[ N_c (2F-1) +F (2F-7) \right] , 
  \nonumber \\ 
C_3 ({\rm S}_{N_c-1}) & = &   \frac{1}{4F^2} 
            (N_c-1) (N_c+2F-1) (N_c+F-1) (2F-1) (F-1) , 
  \nonumber \\ 
C_3 ({\rm S}_1) & = & \frac{1}{4F^2} (F^2-1) (4F^2-1) .
\end{eqnarray}

Regarding the last source of operator identities for the
symmetric case, the statement for the mixed-symmetry case is
simple.  All representations that appear in a product of two
one-body operators also appear in the product $\overline{\rm MS}
\otimes {\rm MS}$.  No additional operator identities follow.

For the mixed-symmetry representation at the three-body level,
there are two large representations (called $\bar b b_0$ and
$adj_3$ in Appendix~\ref{sec:apb}) that annihilate the baryon
states.  However, one can show that the operators in our list
that could have overlap with these representations are all
independent when acting on the physical baryon states.  Thus,
no further operator reduction rules occur for the flavor-singlet
mass operators.

        The summary of operator reduction rules for the
mixed-symmetry representation nonstrange baryons therefore
reads as follows: Decompose the mixed-symmetry generators into
sums of separate core and excited quark pieces as labeled
above.  One may apply the operator reduction rules of
\cite{DJM2} to the core generators alone, and one
may also eliminate $gG_c$.

\section{Counting Operators}

        The building blocks from which one forms operators relevant to
$\ell = 1$ baryons consists of the core operators $S_c^i$, $T_c^a$,
and $G_c^{ia}$, the excited quark operators $s^i$, $t^a$, and $g^{ia}$, and
the orbital angular momentum operator $\ell^i$.  The mixed-symmetry 
representation baryons have orbital quantum number $\ell=1$, and therefore 
the only required combinations of $\ell^i$ are $\openone$ ($\Delta \ell = 0$), 
$\ell^i$ ($\Delta \ell = 1)$, and the $\Delta \ell = 2$ tensor
\begin{equation} \label{l2}
\ell^{(2) \, ij} \equiv \frac 1 2 \{ \ell^i, \ell^j \} -
\frac{\ell^2}{3} \delta^{ij} .
\end{equation}
Since the physical $N_c=3$ baryons have only two core valence
quarks, one needs only consider operators that involve up to two
core quarks in the large $N_c$ analysis.  The operator reduction rules of
\cite{DJM2} state that one may eliminate all core contractions on
flavor indices using $\delta^{ab}$, $d^{abc}$, or $f^{abc}$, or on
spin indices in two $G_c$'s using $\delta^{ij}$ or $\epsilon^{ijk}$.

        We construct in this paper the complete set of time-reversal
even, rotationally-invariant, isosinglet operators for the nonstrange
excited baryons.  There are $18$ such independent operators.  (Incidentally,
for three flavors there are $20$ operators.  The difference between
the two cases is that for two flavors one has an additional operator
reduction
\begin{equation}
S_c^i G_c^{ia} = \frac{1}{4} (N_c+1) T_c^a .
\end{equation}
For more than two flavors, the operators $t S_c G_c$ and 
$\ell^i g^{ia} S_c^j G_c^{ja}$ must be included.)  For the $18$
operators surviving for two flavors the explicit power of
$N_c$ for a given operator is determined by using the large
$N_c$ counting given in Section II.  Factors of $1/N_c^{n-1}$
are included in the definition of the operators, as can be
seen in Table~I.  The full large $N_c$ counting of the matrix 
elements is $O(N_c^{1-n+m})$, where $m$ is the number of times
the coherent operator $G_c^{ia}$ appears. (For more than two flavors,
$T^a_c$ is also potentially coherent.) In Table I we have organized
the operators by the overall order of their matrix elements in the
$1/N_c$ expansion.  Note that the nonstrange {\bf 70}-plet baryons require 
7 masses and 2 mixing angles, so that matrix elements of 9 operators 
of the 18 shown are necessarily linearly dependent upon the other 9 when 
restricted to these states.

        Furthermore, the analysis here is carried out for arbitrary
values of $N_c$, and the matrix elements of a given operator are
usually not homogeneous in $N_c$.  It can happen that matrix elements
of a given set of operators are linearly independent for $N_c = 3$ but
dependent for other values, in particular $N_c \to \infty$.  This
turns out to be the case for $\left< \ell s \right>$ and $\left< \ell
t G_c \right>$, which are both $O(N_c^0)$, but $\left< \ell s +4\ell
tG_c/(N_c+1) \right>$ is $O(1/N_c)$, so that only one of the original
two truly represents an independent $O(N_c^0)$ operator.  This
result is dependent on the particular states (here nonstrange baryons) used
for evaluating the matrix elements.  Since no operator has been eliminated,
such a result is not an operator reduction, but rather what
we call an operator demotion.

        In our analysis of the masses and mixing angles of 
nonstrange baryons, we begin with the leading
operators $N_c \openone$, $\ell s$, and $\ell^{(2)} gG_c /N_c$ (see 
Table~\ref{ops}), which are independent for both $N_c=3$ and $N_c \to \infty$.
We then add subsets of the 8 operators appearing at $O(1/N_c)$ in 
Table~\ref{ops} plus the demoted $O(1/N_c)$ combination 
$\ell s + 4\ell t G_c/(N_c+1)$, in search of a complete set of 9 independent 
operators.  A number of subsets consisting of 6 such $O(1/N_c)$ operators 
complete the basis acting upon the baryon states for $N_c=3$. As one can
show by considering all possibilities, at least one
of these operators is linearly dependent for $N_c \to \infty$.  This
means that one combination of the $O(1/N_c^1)$ operators can be
demoted to $O(1/N_c^2)$.  Using the labels of Table~\ref{matel1}, we 
choose ${\cal O}_9\equiv (N_c+1)/N_c \cdot {\cal O}_4 
+ {\cal O}_5 + 8 \ell^i g^{ja} \{ S_c^j, G_c^{ia}\}/N_c^2 $, which has
$O(1/N_c^2)$ matrix elements.  This gives us an optimal basis, which we
define as a basis where the number of demoted operator combinations is 
maximized\footnote{Beginning with the three leading-order operators, there 
are numerous other choices for the remaining six that provide an operator 
basis that is linearly independent for $N_c=3$ and rank 8 for 
$N_c \rightarrow \infty$. Using the operator definitions in 
Tables~\ref{matel1} and \ref{matel2}, and letting 
${\cal O}_9^\prime \equiv \ell^i g^{ja} \{ S_c^j, G_c^{ia}\}$, one 
can check that all such sets contain ${\cal O}_6$ and ${\cal O}_9^\prime$,
one of ${\cal O}_7$ and ${\cal O}_{11}$, one of ${\cal O}_8$ and 
${\cal O}_{12}$, and two of ${\cal O}_4$, ${\cal O}_5$, and ${\cal O}_{10}$.
An optimal basis can be formed by taking appropriate linear 
combinations.}.

        The set of operators we choose, along with their matrix
elements computed for the nonstrange $\ell=1$ baryon states, is presented in
Table~\ref{matel1}.  This set is identical to that in
Ref.~\cite{CCGL}, except that we replace $tT_c/N_c$ by $sS_c/N_c$,
and $\ell^i g^{ja} \{ S_c^j, G_c^{ia}\}/N_c^2 $ by the demoted operator
defined immediately above.  Table~\ref{matel2} presents, for
completeness, the matrix elements of the remaining 9 operators.

\section{Numerical Analysis}

        The nine mass parameters of the nonstrange $\ell=1$ baryons
appearing at $N_c=3$ consist of diagonal elements of two isospin-3/2
states, $\Delta_{1/2}$ and $\Delta_{3/2}$, and five isospin-1/2
states, $N_{1/2}$, $N^\prime_{1/2}$, $N_{3/2}$, $N^\prime_{3/2}$, and
$N^\prime_{5/2}$; here the subscript indicates total baryon spin,
while total quark spin is indicated by the absence (1/2) or presence
(3/2) of a prime.  To round out the set of mass parameters, observe
that there is one mixing angle for $N^\prime_{1/2}$-$N_{1/2}$ and one
for $N^\prime_{3/2}$-$N_{3/2}$.

        In Ref.~\cite{CCGL} we showed that fits of these nine mass
parameters lead to an unexpected result: Only a few of the
coefficients of the effective Hamiltonian turn out to be of a natural
size (namely, about a few hundred MeV), with the rest being
anomalously small or even consistent with zero.  This analysis was
performed with certain particular sets of operators that did not fully
take into account the demotions described above, and one may wonder
whether these results were a fluke resulting from an unfortunate
choice of basis.  In the current work we possess rules for obtaining
optimal demoted sets of operators as described in the previous
section, and have found that fits using a number of such different
choices lead to similar results.  In particular, with the same mass
eigenvalues and mixing angles as in \cite{CCGL} and the operator basis
listed in Table~\ref{matel1}, one obtains the coefficients $c_i$
defined by the relations
\begin{equation}
M_j = \sum_{i=1}^9 c_i \left< {\cal O}_i \right>_j ,
\end{equation}
where $j = 1, \ldots, 9$ represent mass bilinears, the rows of
Table~\ref{matel1}.  The results of this inversion are presented in
Table~\ref{invert}.  One sees that this fit is nearly identical to
that of Table~III in \cite{CCGL}, in particular that the operators
$\openone$, $\ell^{(2)} gG_c/N_c$, and $S_c^2/N_c$ again appear to be
by far the most significant.  Replacing $tT_c$ by $sS_c$ and using 
the demoted $1/N_c^2$ combination ${\cal O}_9$ has little effect except 
to drastically decrease coefficient uncertainties in some cases.

        The chief implication of Table~IV is to reinforce confidence
in the fits given in Ref.~\cite{CCGL}: The operators chosen were not
completely optimal, but nevertheless represent the optimal choice
quite well.  Moreover, the operators used in the other fits (Tables
III, IV, V) in \cite{CCGL}, are the same as in the basis used here,
and therefore direct comparisons between those fits and
this work are immediate.

        In fact, the only other fits we wish to present here are those
in which the mixing angles are neither taken from pion decays
\cite{CGKM} nor photoproduction data \cite{CC}, but rather make use
only of the seven measured mass eigenvalues and predict the mixing
angles.  Tables \ref{lowest}, \ref{6param}, and \ref{3param} are the
analogues to Tables III, IV, and V in \cite{CCGL}, respectively.  Note
particularly the following features: In Table~\ref{lowest}, it is
again seen that the three leading operators ${\cal O}_1$, ${\cal
O}_2$, ${\cal O}_3$ in the $1/N_c$ (orders $N_c^1$ and $N_c^0$ only)
give a poor accounting for the data, even when including only mass
eigenvalues; furthermore, the predicted mixing angles are nowhere near
the experimental values from \cite{CGKM} or \cite{CC}.  This is no
surprise, since one expects the next order corrections to be of the 
same order as the mass splittings. Indeed, when
three additional operators ${\cal O}_{4,5,6}$, with matrix elements of
$O(1/N_c)$, are included (Table~\ref{6param}), the situation becomes
much better: In addition to an excellent $\chi^2$/d.o.f.\ of 0.23, one
finds that the mixing angles predicted from a mass analysis  naturally
approach the values obtained from decays.  Nevertheless, only ${\cal
O}_1$, ${\cal O}_3$, and ${\cal O}_6$ appear significant; what if one
performs a fit using only those three operators?  The answer is in
Table~\ref{3param}.  Here the results are most surprising: Now the
operator ${\cal O}_3$ actually adjusts its coefficient to give a {\em
small\/} contribution; the $\chi^2$/d.o.f.\ = 0.73 is not bad, but
while the prediction for the spin-3/2 angle is excellent, the
prediction for the spin-1/2 angle is off by about 2$\sigma$.  Even
though ${\cal O}_3$ now looks insignificant, it is actually required
to give nonzero values to the mixing angles, for observe from
Table~\ref{matel1} that ${\cal O}_1$ and ${\cal O}_6$ do not
contribute to mixing.

Also, neither ${\cal O}_3$ nor ${\cal O}_6$ contribute to the
mass splitting $\Delta_{3/2}-\Delta_{1/2}$.  Among the $O(1/N_c)$ or 
larger operators, only the spin-orbit terms split the $\Delta_J$.  In fact, 
the main effect of the spin-orbit terms is to split the $\Delta_J$ states; 
they also contribute to the nucleon mixing, but their effect on the nucleon 
masses is slight because of cancellations. (The coefficients of the two 
spin-orbit terms have opposite signs, unlike what would be expected from a
single overall spin-orbit term $\ell\cdot S= \ell\cdot s + \ell \cdot S_c$.)
So while the spin-orbit terms are small compared to $1/N_c$ expectations,
they do have some importance and one may expect that the errors on the two
coefficients are correlated.  Reduced experimental error bars on the 
$\Delta_J$ states would clarify the role of and need for the spin-orbit
terms.

We conclude from the results presented here and in Ref.~\cite{CCGL} that 
the large $N_c$ operator analysis reproduces both the experimentally measured 
masses and the mixing angles extracted from the strong
and electromagnetic decays.  We have shown here that fits to
the mass eigenvalues alone may be used to predict these angles 
successfully, and have found that this result holds, to varying degree, in 
both six and three operator fits.  These fits reveal that the
$\chi^2$ function is shallow with respect to the mixing angles, 
so that a small $\chi^2$ is obtained in Table~\ref{3param} using only three
operators, even when the mixing angle predictions begin to diverge from the 
decay analyses results.  Our conclusions are unaffected by our 
choice of operator basis, which differs from that of Ref.~\cite{CCGL}.

\section{Vintage SU(6) Analyses}

Operator analyses of baryon masses were performed long before the
$1/N_c$ expansion was proposed.  A main difference between 
modern work and the older work is that one can estimate the
importance of each operator by the order in $1/N_c$ at which it
contributes to the mass.  Inevitably, there are other
differences as well.  In this section, we contrast what we
have done with some of the early work.

Greenberg and Resnikoff~\cite{gr67} (GR) led the way in
performing an analysis based on SU(6), and were later joined by
Horgan and Dalitz~\cite{hd73} (HD). Additionally, there was
work on numerical fits to the baryon mass spectrum
separately from those papers that laid out the operators.  At a
minimum in this context, we should mention the work in 
Refs.~\cite{ctc,jll74,h74,jdh77}. The last of these papers also
corrected some small (as it turned out) numerical errors in the
previous analyses.  All the analyses make the assumption that
only one- and two-body operators enter.  For the nonstrange
members of the {\bf 70}-plet, the early analyses found 5 independent
operators, three of which are independent of the orbital angular
momentum $\ell$, one linear in $\ell$, and one quadratic in
$\ell$.  Five operators is many fewer than we use.  We
need to explain how the differences come about.  We will use
the notation of  GR and give a brief reprise of their
logic. 

Note before starting that GR and HD use wavefunctions that involve
only relative position coordinates, whereas we use Hartree or
independent particle wavefunctions, that is, wavefunctions
relative to a fixed center of mass.  The Hartree wave functions
are exact in the $N_c \rightarrow \infty$ limit.  This leads to
some difference in reckoning what is a one-body, two-body, or
three-body operator.  For example, we consider $\ell$ and
$\ell\cdot s$ as one body operators.  The equivalent in
GR or HD would be a sum over quarks $\alpha$ of
$L_\alpha \cdot \sigma_\alpha$, where $L_\alpha$ is
interpreted~\cite{hd73} as the orbital angular momentum of one
quark with repect to the center-of-mass of the others.  They
would consider this a three body operator, and do not use it. 
The differences between the older work and the present work due
to this point of counting are least apparent in operators with
no factors of angular momentum, and we turn first to them.

For one body terms, one needs operators that have matrix
elements between the {${\bf 6}$} and $\bar {\bf 6}$ of SU(6), and one knows
that

\begin{equation}
{\bf 6} \otimes \bar {\bf 6} = {\bf 1} \oplus {\bf 35}  .
\end{equation}

\noindent Looking for suitable spin-0 operators on the right
hand side, there is only the $T_1^1$.  The notation is 
$T_{{\rm dim SU(6)}}^{{\rm dim SU(3)}}$, and we only consider SU(3)
singlets since we are considering neither strangeness nor isospin breaking.  
For two-body operators, we first note that 

\begin{equation}
{\bf 6} \otimes {\bf 6} = {\bf 15} \oplus {\bf 21},
\end{equation}

\noindent where the ${\bf 15}$ is antisymmetric and the ${\bf 21}$ is
symmetric.  Then we examine the product

\begin{equation}
{\bf 21} \otimes \overline{{\bf 21}} = {\bf 1} \oplus {\bf 35} \oplus {\bf 405}
\end{equation}

\noindent for suitable operators, finding another $T_1^1$ and a
$T_{405}^1$.  Similarly, the product

\begin{equation}
{\bf 15} \otimes \overline{{\bf 15}} = {\bf 1} \oplus {\bf 35} \oplus {\bf 189}
\end{equation}

\noindent yields still one more $T_1^1$ and a $T_{189}^1$.  All
the $T_1^1$'s give equivalent results for a given multiplet, so
we are left with 3 independent spin-0 mass operator candidates,
namely $T_1^1$, $T_{189}^1$, and $T_{405}^1$.

On our list, we have four one- and two-body operators that
contain no orbital angular momentum. They are

\begin{equation}
\openone, S_c^2, \ s \cdot S_c,  \ {\rm \ and\ } t \cdot T_c  \ .
\end{equation}

\noindent From our viewpoint, there is a further, tacit, assumption made 
by the earlier authors~\cite{gr67,hd73}:  Their two-body operators do not
distinguish between, in our language, two S-wave quarks and
an S-wave/P-wave pair.   This implies that whatever physics leads to the 
$\sigma_i \cdot \sigma_j$ terms in the effective mass operator would give
the same coefficient for any pair of quarks, whatever their wave
functions.  If so, the coefficients of our $S_c^2$ and $s \cdot S_c$ terms 
would not be independent, and we would have the same number of independent 
spin-0 operators as GR or HD.  Indeed, with explicit matrix elements 
given by GR, we can verify the linear dependence of our operators upon
theirs or vice-versa.

Next we look for spin-1 operators that can be dotted into the
orbital angular momentum to give rotationally invariant operators.
 GR and HD only consider angular momentum which is the relative angular 
momentum of a quark pair.  Since unit angular momentum requires antisymmetry, 
GR and HD use only the antisymmetric [in SU(6)] ${\bf 15}$ two-quark 
combination, and find only the operator $T_{35}^1$.  Perusing our list, we 
find 3 operators at the one- or two-body level that use $\ell$ once:

\begin{equation} \label{this}
\ell \cdot s, \ \ell \cdot S_c, \ {\rm \ and\ } 
       \ell g T_c  .
\end{equation}

\noindent The question of how to connect our $\ell$ to their
angular momentum operator returns.  If $\ell$ is the angular
momentum of one quark relative to the overall center of mass,
it is a three-body operator, as discussed earlier, and thus
would be discarded by the early authors.  For us, $\ell$ is the
angular momentum with respect to the center of mass, and we can
interpret part of it as the angular momentum of the excited quark
with respect to one particular core quark.  Then, matching to the
earlier authors, $\ell \cdot s$ and $\ell \cdot S_c$ would have the 
same coefficient by arguments already made, if $S_c$ is taken to refer to a 
quark in that pair (and if not, it would be a three-body term).  Regarding our
third term, again following GR or HD, we would apply it only to antisymmetric 
subsets of quarks, and for either purely symmetric~\cite{DJM2} or purely 
antisymmetric quark states one can prove a result [a consequence of
Eq.~(\ref{cubic})] that

\begin{equation}
g^{ia} T^a_c \propto S^i_c+s^i  ,
\end{equation}

\noindent  Thus, the third spin-1 operator in (\ref{this}) becomes dependent 
upon the first two.

Similarly, spin-2 operators that can be combined with the 
$\Delta \ell = 2$ part of $\ell^i \ell^j$ come from the symmetric {\bf 21}
in the earlier authors's analysis.  Here, they find only an operator
$T_{405}^1$.  We have two operators at the two- or fewer-body
level, which are

\begin{equation}
\ell^{(2)} g G_c {\rm \ and\ }
    \ell^{(2)} s S_c  \ .
\end{equation}

\noindent But again, if we ignore differences between quarks and
recall that GR or HD would only let the operators act on
symmetric states, there are operator reduction rules 
stating\footnote{To be explicit, this is the third identity from
the bottom of Table~VI in \cite{DJM2}.}

\begin{equation}
g^{ia} G^{ja}_c \propto s^i S^j_c
\end{equation}

\noindent for the spin-2 piece, and again only the core quark that appears
in the pair under discussion is meant above.  Hence in this view, we 
would have one operator also.

Thus if we make GR's or HD's assumptions, we get their results. 
However, our analysis is more general and relies only on an
organizing principle suggested by the underlying theory.
On the practical side, GR did not use the tensor
operator in their fits, on the grounds that there was not 
enough data at that time to justify one  more operator.  We found that this
operator was quite important.  They did, however, find that the
spin-orbit operators had small coefficients~\cite{h74}, a result
that was confirmed by Isgur and Karl~\cite{ik78}.

\section{Dynamical Interpretation}

The most striking feature of our analysis is that the nonstrange $\ell=1$
mass spectrum is described adequately by two nontrivial operators,
\begin{equation}
\frac{1}{N_c} \ell^{(2)} g \,G_c \,\,\,\,\, \mbox{ and } 
\,\,\,\,\,
\frac{1}{N_c} S_c^2 \,\,\, .
\label{eq:twops}
\end{equation}
Clearly, large $N_c$ power counting is not sufficient by itself to
explain the $\ell=1$ baryon masses---the underlying dynamics
plays a crucial role.  In this section, we simply point out that the
preferred set of operators in Eq.~(\ref{eq:twops}) can be understood
in a constituent quark model with a single pseudoscalar meson
exchange, up to corrections of order $1/N_c^2$.  The argument goes as 
follows:

The pion couples to the quark axial-vector current so that the 
$\overline{q}q\pi$ coupling introduces the spin-flavor structure
$\sigma^i \tau^a$ on a given quark line. In addition, pion exchange 
respects the large $N_c$ counting rules given in Section~II.  A single 
pion exchange between the excited quark and a core quark corresponds
to the operators
\begin{equation}
g^{ia} G_c^{ja} \ell^{(2)}_{ij} 
\label{eq:opone}
\end{equation}
and
\begin{equation}
g^{ia} G_c^{ia}
\label{eq:optwo}
\end{equation}
while pion exchange between two core quarks yields 
\begin{equation}
G^{ia}_c G^{ia}_c \,\,\, .
\label{eq:opthree}
\end{equation}
These exhaust the possible two-body operators that have the desired 
spin-flavor structure (since $\ell^{(2)} G_c G_c$ is a three-body
operator).  The first operator is one of the two in our 
preferred set. The third operator may be rewritten 
\begin{equation}
2 G^{ia}_c G^{ia}_c = C_2 \cdot \openone - \frac{1}{2}T^a_c T^a_c 
- \frac{1}{2} S_c^2
\label{eq:cas}
\end{equation}
where $C_2$ is the SU(4) quadratic Casimir for the totally symmetric core
representation (the {\bf 10}  of SU(4) for $N_c=3$).  Since the core 
wavefunction involves two spin and two flavor degrees of freedom, and is 
totally symmetric, it is straightforward to show that  $T_c^2=S_c^2$.  Then
Eq.~(\ref{eq:cas}) implies that one may exchange $G^{ia}_c G^{ia}_c$ in 
favor of the identity operator and $S_c^2$, the second of the two operators 
suggested by our fits.

The remaining operator, $g^{ia} G_c^{ia}$, is peculiar in that
its matrix element between two nonstrange, mixed symmetry states is given 
by 
\begin{equation}
\frac{1}{N_c} \langle gG \rangle = - \frac{N_c+1}{16 N_c} +
\delta_{S,I}\frac{I(I+1)}{2N_c^2}  \,\,\, ,
\end{equation}
which differs from the identity only at order $1/N_c^2$.  Thus to order 
$1/N_c$, one may make the
replacements
\begin{equation}
\{\openone \mbox{ , } g^{ia} G_c^{ja} \ell^{(2)}_{ij}  \mbox{ , }
g^{ia} G_c^{ia} \mbox{ , } G^{ia}_c G^{ia}_c \} \Rightarrow
\{\openone  \mbox{ , }  g^{ia} G_c^{ja} \ell^{(2)}_{ij} \mbox{ , }
S_c^2\} \,\,\, .
\end{equation}
We conclude that the operator set suggested by the data may
be understood in terms of single pion exchange between quark
lines.  This is consistent with the interpretation of
the mass spectrum advocated by Glozman and Riska~\cite{gloz}.
Other simple models, such as single gluon exchange, do not
directly select the operators suggested by our analysis and
may require others that are disfavored by the data.

\section{Summary and Conclusions}

We have considered what the large $N_c$ expansion tells about
the masses of the non-strange P-wave excited baryons.  We have
given the effective mass operator by enumerating all the independent
operators that it could contain, and ordered those operators by
their size in the $1/N_c$ expansion.  We have calculated the
matrix elements of each of the operators for any $N_c$.  For the effective 
mass operator, we have fit the coefficients of the individual operators 
to the data, using the masses given by the Particle Data
Group~\cite{pdg} and after truncating the full set of operators
in suitable and reasonable ways.

We find that one can fit the masses well using selected subsets
of the full list of operators,  and that the good fits have
mixing angles that are compatible with the mixing angles
that come from analyses of the mesonic and radiative decays of
these baryons~\cite{CGKM,CC,workman}.  Estimating the size of
each operator using the $1/N_c$ scheme works, in the sense that
no operator is larger than expected based on those estimates. 
Some operators are smaller.  In fact, we can get a decent fit
keeping just the unit operator, one tensor operator, and the
core spin-squared operator.   This is compatible with the idea
that the underlying dynamics is due to effective pseudoscalar meson
exchanges among the quarks~\cite{gloz}, and not easily
compatible with the idea that the masses splittings are
explained by single gluon exchange. 

{\samepage
\begin{center}
{\bf Acknowledgments}
\end{center}
CEC thanks the National Science Foundation for support under Grant
PHY-9600415; CDC thanks the NSF for support under Grant PHY-9800741;
and JLG similarly thanks the NSF for support under Grant HRD-9633750.
Both JLG and RFL thank the Department of Energy for support under
Contract DE-AC05-84ER40150.  In addition, RFL thanks the Institute for
Nuclear Theory at the University of Washington and the University of
California, San Diego for their hospitality during the preparation of
the manuscript, and Aneesh Manohar for comments on the fits.}

\appendix
\section{Explicit matrix elements}\label{sec:apa}
Using the notation for quantum numbers defined in Eq.~(\ref{states}),
we first present matrix elements of the SU(6) generators between
completely symmetric core states:
\begin{eqnarray}
  \left< S_c^\prime = I_c^\prime; m_1^\prime, \alpha_1^\prime \left|
      G_c^{ia} \right| S_c = I_c; m_1, \alpha_1 \right> & = & \frac 1
  4 \sqrt{\frac{2I_c+1}{2I_c^\prime+1}} \sqrt{(N_c+1)^2 - (I_c^\prime
    - I_c)^2 (I_c^\prime + I_c + 1)^2} \nonumber \\ & & \times \left(
    \begin{array}{cc} S_c & 1 \\ m_1 & i \end{array} \right| \left.
    \begin{array}{c} S^\prime_c \\ m_1^\prime \end{array} \right) \left(
    \begin{array}{cc} I_c & 1 \\ \alpha_1 & a \end{array} \right| \left.
    \begin{array}{c} I^\prime_c \\ \alpha_1^\prime \end{array} \right)
\label{eq:giame}
  , \\ \left< S_c^\prime = I_c^\prime; m_1^\prime, \alpha_1^\prime \left|
      T_c^a \right| S_c = I_c; m_1, \alpha_1 \right> & = & \sqrt{I_c
    (I_c+1)} \left(
    \begin{array}{cc} I_c & 1 \\ \alpha_1 & a \end{array} \right| \left.
    \begin{array}{c} I_c \\ \alpha_1^\prime \end{array} \right)
  \delta_{I_c^\prime I_c} \delta_{S_c^\prime S_c} \delta_{m_1^\prime
    m_1} , \\ \left< S_c^\prime = I_c^\prime; m_1^\prime,
    \alpha_1^\prime \left| S_c^i \right| S_c = I_c; m_1, \alpha_1
  \right> & = & \sqrt{I_c (I_c+1)} \left(
    \begin{array}{cc} S_c & 1 \\ m_1 & i \end{array} \right| \left.
    \begin{array}{c} S_c \\ m_1^\prime \end{array} \right)
  \delta_{I_c^\prime I_c} \delta_{S_c^\prime S_c}
  \delta_{\alpha_1^\prime \alpha_1} .
\end{eqnarray}
To obtain the matrix elements of $s, t, g$ in terms of those for $S_c,
T_c, G_c$, simply note that the excited quark is group-theoretically
equivalent to a one-quark core with spin and isospin 1/2.  Thus,
replace each $N_c-1$ by 1, and each $S_c=I_c$ and $S_c^\prime =
I_c^\prime$ by 1/2.  The matrix elements of the orbital angular 
momentum operators are
\begin{eqnarray}
\left< \ell^\prime
    m^\prime_\ell \left| \ell^i \right| \ell m_\ell \right> & = &
  \sqrt{\ell(\ell+1)} \left( \begin{array}{cc} \ell & 1 \\ m_\ell 
& i \end{array}
  \right| \left. \begin{array}{c} \ell \\ m_\ell^\prime \end{array}
  \right) \delta_{\ell^\prime \ell} , \\ \left< \ell^\prime
    m^\prime_\ell \left| \ell^{(2) \ ij} \right| \ell m_\ell \right> &
  = & \sqrt{\frac{\ell (\ell+1) (2\ell-1) (2\ell+3)}{6}}
  \delta_{\ell^\prime \ell} \nonumber \\ & & \times \sum_\mu \left(
  \begin{array}{cc} 1 & 1 \\ i & j \end{array} \right| \left.
  \begin{array}{c} 2 \\ \mu \end{array} \right)\left(
  \begin{array}{cc} \ell & 2 \\ m_\ell & \mu \end{array} \right| \left.
  \begin{array}{c} \ell \\ m_\ell^\prime \end{array} \right) .
\end{eqnarray}
With these results we have computed the matrix elements of all the
possible isosinglet mass operators:
\begin{equation}
\left< \openone \right> = \delta_{J^\prime J} \delta_{J^\prime_3 J_3}
\delta_{\ell^\prime \ell} \delta_{I^\prime I} \delta_{I^\prime_3 I_3}
\delta_{S^\prime S} ,
\end{equation}
\begin{eqnarray}
\left< \ell s \right> & = & \delta_{J^\prime J} \delta_{J^\prime_3
J_3} \delta_{\ell^\prime \ell} \delta_{I^\prime I} \delta_{I^\prime_3
I_3} (-1)^{S^\prime-S} \, \frac 1 2 \sqrt{\left( 2S^\prime + 1 \right)
\left( 2S + 1 \right)} \nonumber \\
& \times & \sum_{\ell_s = \ell \pm 1/2} \left[ \ell_s \left( \ell_s +
1 \right) - \ell \left( \ell + 1 \right) - \frac 3 4 \right] \left( 2
\ell_s + 1 \right) \nonumber \\
& \times & \sum_{\eta = \pm 1} c_{\rho^\prime \eta} c_{\rho \eta}
\left\{ \begin{array}{ccc} I_c & \frac 1 2 & S^\prime \\ \ell & J &
\ell_s \end{array} \right\} \left\{ \begin{array}{ccc} I_c & \frac 1 2
& S \\ \ell & J & \ell_s \end{array} \right\} ,
\end{eqnarray}
\begin{eqnarray}
\left< \ell t G_c \right> & = & \delta_{J^\prime J} \delta_{J^\prime_3
J_3} \delta_{\ell^\prime \ell} \delta_{I^\prime I} \delta_{I^\prime_3
I_3} (-1)^{J+I+\ell + S^\prime - S +1} \, \frac 1 4 \sqrt{\frac 3 2}
\sqrt{\ell (\ell+1) (2\ell + 1)} \sqrt{(2S^\prime + 1)(2S + 1)}
\nonumber \\
& \times & \sum_{\eta^\prime, \eta = \pm 1} c_{\rho^\prime \eta}
c_{\rho \eta} \sqrt{ (2I^\prime_c + 1) (2I_c + 1)} \sqrt{(N_c+1)^2 -
\left( \frac{\eta^\prime - \eta}{2} \right)^2 (2I+1)^2} \nonumber \\
& \times & \left\{ \begin{array}{ccc} \frac 1 2 & 1 & \frac 1 2 \\
I_c^\prime & I & I_c \end{array} \right \} \left\{ \begin{array}{ccc}
\ell & 1 & \ell \\ S^\prime & J & S \end{array} \right\} \left\{
\begin{array}{ccc} S^\prime & 1 & S \\ I_c & \frac 1 2 & I_c^\prime
\end{array} \right\} (-1)^{(\eta^\prime - \eta)/2} ,
\end{eqnarray}
\begin{eqnarray}
\left< \ell^{(2)} gG_c \right> & = & \delta_{J^\prime J}
\delta_{J^\prime_3 J_3} \delta_{\ell^\prime \ell} \delta_{I^\prime I}
\delta_{I^\prime_3 I_3} (-1)^{J-2I+\ell+S} \, \frac 1 8
\sqrt{\frac{15}{2}} \sqrt{\ell (\ell+1) (2\ell-1) (2\ell+1) (2\ell+3)}
\nonumber \\
& \times & \sqrt{(2S^\prime + 1)(2S+1)} \left\{ \begin{array}{ccc} 2 &
\ell & \ell \\ J & S^\prime & S \end{array} \right\} \nonumber \\
& \times & \sum_{\eta^\prime, \eta = \pm 1} c_{\rho^\prime
\eta^\prime} c_{\rho \eta} (-1)^{(1 + \eta^\prime)/2}
\sqrt{(2I_c^\prime + 1)(2I_c +1)} \sqrt{(N_c+1)^2 - \left(
\frac{\eta^\prime - \eta}{2} \right)^2 (2I+1)^2} \nonumber \\
& \times & \left\{ \begin{array}{ccc} \frac 1 2 & 1 & \frac 1 2 \\
I_c^\prime & I & I_c \end{array} \right\} \left\{ \begin{array}{ccc}
I_c^\prime & I_c & 1 \\ S^\prime & S & 2 \\ \frac 1 2 & \frac 1 2 & 1
\end{array} \right\} ,
\end{eqnarray}
\begin{eqnarray}
\left< \ell S_c \right> & = & \delta_{J^\prime J} \delta_{J^\prime_3
J_3} \delta_{\ell^\prime \ell} \delta_{I^\prime I} \delta_{I^\prime_3
I_3} (-1)^{J-I+\ell+S^\prime-S} \sqrt{\ell (\ell+1) (2\ell+1)}
\sqrt{(2S^\prime + 1) (2S+1)} \nonumber \\
& \times & \left\{ \begin{array}{ccc} 1 & \ell & \ell \\ J & S^\prime
& S \end{array} \right\} \sum_{\eta = \pm 1} c_{\rho^\prime \eta}
c_{\rho \eta} (-1)^{(1-\eta)/2} \sqrt{I_c (I_c+1)(2I_c+1)} \left\{
\begin{array}{ccc} 1 & I_c & I_c \\ \frac 1 2 & S^\prime & S \end{array}
\right\} ,
\end{eqnarray}
\begin{equation}
\left< tT_c \right> = \delta_{J^\prime J} \delta_{J^\prime_3 J_3}
\delta_{\ell^\prime \ell} \delta_{I^\prime I} \delta_{I^\prime_3 I_3}
\delta_{S^\prime S} \, \frac 1 4 \left[ 4I(I+1) - 3 - 4 \sum_{\eta =
\pm 1} c_{\rho \eta}^2 I_c (I_c + 1) \right] ,
\end{equation}
\begin{equation}
\left< S_c^2 \right> = \delta_{J^\prime J} \delta_{J^\prime_3 J_3}
\delta_{\ell^\prime \ell} \delta_{I^\prime I} \delta_{I^\prime_3 I_3}
\delta_{S^\prime S} \sum_{\eta = \pm 1} c_{\rho \eta}^2 I_c (I_c+1) ,
\end{equation}
\begin{eqnarray}
\left< \ell g T_c \right> & = & \delta_{J^\prime J} \delta_{J^\prime_3
J_3} \delta_{\ell^\prime \ell} \delta_{I^\prime I} \delta_{I^\prime_3
I_3} (-1)^{J+I+\ell+ 1} \, \frac 3 2 \sqrt{\ell (\ell+1) (2\ell+1)}
\sqrt{(2S^\prime+1) (2S+1)} \nonumber \\
& \times & \left\{ \begin{array}{ccc} 1 & \ell & \ell \\ J & S^\prime
& S \end{array} \right\} \sum_{\eta = \pm 1} c_{\rho^\prime \eta}
c_{\rho \eta} \sqrt{I_c (I_c+1) (2I_c+1)} \left\{ \begin{array}{ccc} 1
& \frac 1 2 & \frac 1 2 \\ I & I_c & I_c \end{array} \right\} \left\{
\begin{array}{ccc} 1 & \frac 1 2 & \frac 1 2 \\ I_c & S^\prime & S
\end{array} \right\} ,
\end{eqnarray}
\begin{eqnarray}
\left< \ell^{(2)} s S_c \right> & = & \delta_{J^\prime J}
\delta_{J^\prime_3 J_3} \delta_{\ell^\prime \ell} \delta_{I^\prime I}
\delta_{I^\prime_3 I_3} (-1)^{J + \ell +S} \, \frac{\sqrt{5}}{2}
\nonumber \\
& \times & \sqrt{\ell (\ell+1) (2\ell-1) (2\ell+1) (2\ell+3)}
\sqrt{(2S^\prime+1) (2S+1)} \left\{ \begin{array}{ccc} 2 & \ell & \ell
\\ J & S^\prime & S \end{array} \right\} \nonumber \\
& \times & \sum_{\eta = \pm 1} c_{\rho^\prime \eta}
c_{\rho \eta} \sqrt{I_c (I_c + 1) (2I_c+1)} \left\{ \begin{array}{ccc}
I_c & I_c & 1 \\ S^\prime & S & 2 \\ \frac 1 2 & \frac 1 2 & 1
\end{array} \right\} ,
\end{eqnarray}
\begin{eqnarray}
\left< \ell^i g^{ja} \{ S_c^j, G_c^{ia} \} \right> & = & \delta_{J^\prime J}
\delta_{J^\prime_3 J_3} \delta_{\ell^\prime \ell} \delta_{I^\prime I}
\delta_{I^\prime_3 I_3} (-1)^{J-I+\ell} \, \frac{3}{8} \sqrt{\ell
(\ell+1) (2\ell+1)} \sqrt{(2S^\prime + 1)(2S+1)} \nonumber \\
& \times & \left\{ \begin{array}{ccc} 1 & \ell & \ell \\ J & S^\prime
& S \end{array} \right\} \sum_{\eta^\prime, \eta = \pm 1}
c_{\rho^\prime \eta^\prime} c_{\rho \eta} \left\{ \begin{array}{ccc}
\frac 1 2 & 1 & \frac 1 2 \\ I_c^\prime & I & I_c \end{array} \right\}
\left\{ \begin{array}{ccc} I_c^\prime & 1 & I_c \\ S & \frac 1 2 &
S^\prime \end{array} \right\} \sqrt{(2I_c^\prime+1)(2I_c+1)} \nonumber
\\
& \times & \sqrt{(N_c+1)^2 - \left( \frac{\eta^\prime - \eta}{2}
\right)^2 (2I+1)^2} \nonumber \\
& \times & \left[ (-1)^{S-I_c+1/2} \left\{ \begin{array}{ccc}
\frac 1 2 & 1 & \frac 1 2 \\ I_c^\prime & S^\prime & I_c^\prime
\end{array} \right\} \sqrt{I_c^\prime (I_c^\prime +1)(2I_c^\prime +1)}
\right. \nonumber \\
& & + \left. (-1)^{S^\prime -I_c^\prime +1/2} \left\{
\begin{array}{ccc} \frac 1 2 & 1 & \frac 1 2 \\ I_c & S & I_c
\end{array} \right\} \sqrt{I_c (I_c +1)(2I_c +1)} \right] ,
\end{eqnarray}
\begin{eqnarray}
\left< \ell^{(2)} t \{ S_c, G_c \} \right> & = & \delta_{J^\prime J}
\delta_{J^\prime_3 J_3} \delta_{\ell^\prime \ell} \delta_{I^\prime I}
\delta_{I^\prime_3 I_3} (-1)^{J-I+\ell+S^\prime - S +1} \, \frac 1 8
\sqrt{5\ell (\ell+1) (2\ell-1) (2\ell+1) (2\ell+3)}  \nonumber \\
& \times &\sqrt{(2S^\prime +1) (2S+1)} \left\{ \begin{array}{ccc} 2 &
\ell & \ell \\ J & S^\prime & S
\end{array} \right\} \sum_{\eta^\prime, \eta = \pm 1} c_{\rho^\prime
\eta^\prime} c_{\rho \eta} \sqrt{(2I_c^\prime + 1)(2I_c + 1)}
\nonumber \\
& \times & \sqrt{(N_c+1)^2 - \left( \frac{\eta^\prime - \eta}{2}
\right)^2 (2I+1)^2} \left\{ \begin{array}{ccc} \frac 1 2 & 1 & \frac 1
2 \\ I_c^\prime & I & I_c \end{array} \right\} \left\{
\begin{array}{ccc} 2 & I_c^\prime & I_c \\ \frac 1 2 & S^\prime & S
\end{array} \right\} \nonumber \\
& \times & \left[ \sqrt{I_c^\prime (I_c^\prime + 1)(2I_c^\prime + 1)}
\left\{ \begin{array}{ccc} 2 & 1 & 1 \\ I_c^\prime & I_c^\prime & I
\end{array} \right\} + \sqrt{I_c (I_c+1) (2I_c + 1)} \left\{
\begin{array}{ccc} 2 & 1 & 1 \\ I_c & I_c & I_c^\prime \end{array}
\right\} \right] , \nonumber \\ & &
\end{eqnarray}
\begin{equation}
\left< s S_c \right> = \delta_{J^\prime J} \delta_{J^\prime_3 J_3}
\delta_{\ell^\prime \ell} \delta_{I^\prime I} \delta_{I^\prime_3 I_3}
\delta_{S^\prime S} \, \frac 1 2 \left[ \left( S - \frac 1 2 \right)
\left( S + \frac 3 2 \right) - \sum_{\eta = \pm 1} c_{\rho \eta}^2 I_c
(I_c + 1) \right] ,
\end{equation}
\begin{eqnarray}
\left< (\ell S_c) (tT_c) \right> & = & \delta_{J^\prime J}
\delta_{J^\prime_3 J_3} \delta_{\ell^\prime \ell} \delta_{I^\prime I}
\delta_{I^\prime_3 I_3} (-1)^{J-I+\ell+S^\prime -S +1} \, \frac 1 4
\sqrt{\ell (\ell+1) (2\ell+1)} \nonumber
\\
& \times & \sqrt{(2S^\prime +1) (2S+1)} \left\{ \begin{array}{ccc} 1 &
\ell & \ell \\ J & S^\prime & S \end{array} \right\} \nonumber \\
& \times & \sum_{\eta = \pm 1} c_{\rho^\prime \eta} c_{\rho \eta}
\left[ (2I+1) \eta + 2 \right] (-1)^{(1-\eta)/2} \sqrt{I_c (I_c+1)
(2I_c+1)} \left\{ \begin{array}{ccc} 1 & I_c & I_c \\ \frac 1 2 &
S^\prime & S \end{array} \right\} , \nonumber \\
\end{eqnarray}
\begin{eqnarray}
\left< ( \ell s ) S_c^2 \right> & = & \delta_{J^\prime J}
\delta_{J^\prime_3 J_3} \delta_{\ell^\prime \ell} \delta_{I^\prime I}
\delta_{I^\prime_3 I_3} (-1)^{S^\prime - S} \, \frac 1 2
\sqrt{(2S^\prime + 1) (2S+1)} \nonumber \\
& \times & \sum_{\ell_s = \ell \pm 1/2} \left[ \ell_s (\ell_s +1) -
\ell (\ell+1) - \frac 3 4 \right] (2\ell_s + 1) \nonumber \\
& \times & \sum_{\eta = \pm 1} c_{\rho^\prime \eta} c_{\rho \eta}
\left\{ \begin{array}{ccc} I_c & \frac 1 2 & S^\prime \\ \ell & J &
\ell_s \end{array} \right\} \left\{ \begin{array}{ccc} I_c & \frac 1 2
& S \\ \ell & J & \ell_s \end{array} \right\} I_c(I_c+1) ,
\end{eqnarray}
\begin{eqnarray}
\left< \{ \ell S_c, sS_c \} \right> & = & \delta_{J^\prime J}
\delta_{J^\prime_3 J_3} \delta_{\ell^\prime \ell} \delta_{I^\prime I}
\delta_{I^\prime_3 I_3} (-1)^{J-I+\ell+S^\prime-S} \, \sqrt{\ell
(\ell+1) (2\ell+1)} \sqrt{(2S^\prime+1)(2S+1)} \nonumber \\
& \times & \left\{ \begin{array}{ccc} 1 & \ell & \ell \\ J & S^\prime
& S \end{array} \right\} \sum_{\eta = \pm 1} c_{\rho^\prime \eta}
c_{\rho \eta} \left\{ \frac 1 2 \left[ S^\prime (S^\prime + 1) +
S(S+1) \right] - \frac 3 4 - I_c (I_c+1) \right\}
\nonumber \\
& \times & (-1)^{(1-\eta)/2} \sqrt{I_c (I_c+1) (2I_c+1)} \left\{
\begin{array}{ccc} 1 & I_c & I_c \\ \frac 1 2 & S^\prime & S
\end{array} \right\} ,
\end{eqnarray}
\begin{eqnarray}
\left< g S_c T_c \right> & = &  \delta_{J^\prime J}
\delta_{J^\prime_3 J_3} \delta_{\ell^\prime \ell} \delta_{I^\prime I}
\delta_{I^\prime_3 I_3} \delta_{S^\prime S} (-1)^{I-S} \nonumber \\
& \times & \frac 3 2 \sum_{\eta = \pm 1} c_{\rho \eta}^2 I_c (I_c+1)
(2I_c+1) \left\{ \begin{array}{ccc} \frac 1 2 & 1 & \frac 1 2 \\ I_c &
I & I_c \end{array} \right\} \left\{ \begin{array}{ccc} \frac 1 2 & 1
& \frac 1 2 \\ I_c & S & I_c \end{array} \right\} ,
\end{eqnarray}
\begin{eqnarray}
\left< \ell^{(2)} S_c S_c \right> & = & \delta_{J^\prime J}
\delta_{J^\prime_3 J_3} \delta_{\ell^\prime \ell} \delta_{I^\prime I}
\delta_{I^\prime_3 I_3} (-1)^{J+I-\ell +S^\prime - S}
\sqrt{\frac{5\ell (\ell+1) (2\ell-1) (2\ell+1) (2\ell+3)}{6}}
\nonumber \\
& \times & \sqrt{(2S^\prime + 1) (2S+1)} \left\{ \begin{array}{ccc} 2
& \ell & \ell \\ J & S^\prime & S \end{array} \right\} \sum_{\eta =
\pm 1} c_{\rho^\prime \eta} c_{\rho \eta} I_c (I_c +1) (2I_c+1)
(-1)^{(1-\eta)/2} \nonumber \\
& \times & \left\{ \begin{array}{ccc} 2 & I_c & I_c \\ \frac 1 2 &
S^\prime & S \end{array} \right\} \left\{ \begin{array}{ccc} 2 & \ell
& \ell \\ I_c & I_c & I_c \end{array} \right\} ,
\end{eqnarray}
\begin{eqnarray}
\left< \ell^{(2)} g S_c T_c \right> & = & \delta_{J^\prime J}
\delta_{J^\prime_3 J_3} \delta_{\ell^\prime \ell} \delta_{I^\prime I}
\delta_{I^\prime_3 I_3} (-1)^{J+2I+\ell+S} \sqrt{\frac{5\ell (\ell+1)
(2\ell-1) (2\ell+1) (2\ell+3)}{6}} \nonumber \\
& \times & \frac 3 2 \sqrt{(2S^\prime+1)(2S+1)} \left\{
\begin{array}{ccc} 2 & \ell & \ell \\ J & S^\prime & S \end{array}
\right\} \sum_{\eta = \pm 1} c_{\rho^\prime \eta} c_{\rho \eta} I_c
(I_c+1) (2I_c+1) (-1)^{(\eta+1)/2} \nonumber \\
& \times & \left\{ \begin{array}{ccc} \frac 1 2 & 1 & \frac 1 2 \\ I_c
& I & I_c \end{array} \right\} \left\{ \begin{array}{ccc} I_c & I_c &
1 \\ S^\prime & S & 2 \\ \frac 1 2 & \frac 1 2 & 1 \end{array}
\right\} .
\end{eqnarray}

\section{Operator reductions for mixed-symmetry states}\label{sec:apb}

        In order to establish a connection to the operator
reduction rules obtained in \cite{DJM2} for the completely
symmetric ground-state spin-flavor multiplet, let us first
develop a common notation and then review the results of the
derivation presented in the earlier work.

Each irreducible representation of SU($2F$) is denoted by a
Dynkin label, which is a $(2F-1)$-plet $[n_1, n_2,
\ldots , n_{2F-1} ]$ of nonnegative integers $n_r$ describing
the Young diagram of the representation; the number of boxes in
row $r$ (= 1, 2, \ldots, $2F-1$) of the diagram exceeds the
number in row
$r+1$ by $n_r$.  The conjugate of a given representation is
obtained by reversing the order of the integers $n_r$. In this
notation the completely symmetric $N_c$-box representation S is
$[N_c, 0, 0, \ldots, 0 ]$, while the mixed-symmetry $\ell = 1$
baryons fill the representation MS = $[N_c-2, 1, 0, 0, \ldots,
0 ]$.   Since all matrix elements of operators ${\cal O}$
between baryons $B$ transforming according to a given
representation appear through bilinears of the form $\bar B
{\cal O} B$, such operators fill the representations of $\bar B
\otimes B$.  In the case of the ground-state representation,
standard techniques for combining representations show that
this product is
\begin{equation}
  \bar {\rm S} \otimes {\rm S} 
    = [0, 0, 0, \ldots, N_c ] \otimes [N_c, 0, 0, \ldots, 0 ] 
      = \bigoplus_{m=0}^{N_c} [ m, 0, 0, \ldots, m ] ,
\end{equation}
while the mixed-symmetry representation product gives
\begin{eqnarray}
\overline{\rm MS} \otimes {\rm MS} & = & 
[0, 0, \ldots, 0, 1, N_c-2 ] \otimes 
[N_c-2, 1, 0, 0, \ldots, 0 ] 
              \nonumber \\ 
& = & [0, 0, \ldots, 0] \oplus [N_c-1, 0, 0, \ldots , N_c-1] 
              \nonumber \\ & &
\bigoplus_{m=1}^{N_c-2} 2 \, [m, 0, 0, \ldots, 0, m]
\bigoplus_{m=0}^{N_c-2} [m, 1, 0, 0, \ldots, 0, 1, m ] 
              \nonumber \\ &
& \bigoplus_{m=0}^{N_c-3} [m+2, 0, 0, \ldots, 0, 1, m]
\bigoplus_{m=0}^{N_c-3} [m, 1, 0, 0, \ldots , 0, m+2 ] .
\label{msprod}
\end{eqnarray}
It is convenient to give these representations concise names for future
reference.  Label $[m, 0, 0, \ldots, m]$ as $adj_m$, so that
$adj_0$ is the singlet rep, $adj_1$ is the adjoint rep, and
$adj_2$ is called
$\bar s s$ in \cite{DJM2}.  Let $[m, 1, 0, 0, \ldots, 0, 1, m]
\equiv
\bar a a_m$, $[m+2, 0, 0, \ldots, 0, 1, m] \equiv \bar a s_m$,
and
$[m, 1, 0, 0, \ldots, 0, m+2] \equiv \bar s a_m$, so that $\bar
a a_0$, $\bar a s_0$, and $\bar s a_0$ 
are denoted in~\cite{DJM2} as
$\bar a a$, $\bar a s$, and $\bar s a$, respectively.

That the operators~(\ref{basis}) satisfy an SU($2F$) algebra
implies that any string of the one-body operators in~(\ref{basis})
containing a commutator is reducible to a smaller string of
such operators; only anticommutators need be considered.  Since
one-body operators appear in the adjoint representation of 
SU($2F$), one need only consider combinations symmetrized on
the adjoint indices, $\openone$,
$adj$, $\left( adj \otimes adj \right)_S$, $\left( adj \otimes
adj
\otimes adj \right)_S$, {\it etc.}, which may be denoted 0-,
1-, 2-, {\it etc.}  body operators.

Let us turn to the question of representations that appear in
symmetrized products of one-body operators, but have vanishing 
matrix elements for the mixed-symmetry baryon states
since they do not appear in the product $\overline{\rm MS} \otimes {\rm MS}$.
First note that, unlike $\bar {\rm S} \otimes {\rm S}$, all 
representations in
\begin{equation}
( adj \otimes adj )_S  = adj_0 \oplus adj_1 \oplus adj_2 \oplus
\bar a a_0 ,
\end{equation}
in particular $\bar a a_0$, appear in the product
$\overline{\rm MS}
\otimes {\rm MS}$, meaning that the mixed-symmetry
representation has no similar operator identity.  One therefore
turns to representations in the product
$( adj \otimes adj \otimes adj )_S$, which are listed in
Table~IV of \cite{DJM2}. Comparing this list to (\ref{msprod}),
one sees that the only representations not present in the
latter are $[0, 0, 1, 0, 0, \ldots, 0, 1, 0, 0] \equiv
\bar b b_0$ and $adj_3$.  Products of three one-body operators
that transform according to these representations should indeed
be reducible when acting on mixed-symmetry baryon states. The
astute reader may notice that each of these representations has
a special feature: $\bar b b_0$ does not occur for $F < 3$, and
$adj_3$ only gives reduction rules for the physical case $N_c =
3$. The product of three one-body operators is enough to span the space of 
all physical baryon observables; for this reason, we do not consider
representations in the product of four or more one-body
operators.

\begin{table}
\begin{tabular}{cc|l}
\multicolumn{1}{c}{Order of matrix element} &&
\multicolumn{1}{l}{Operator} \\ \hline\hline

$N_c^1$ && $N_c \openone$ \\ \hline

$N_c^0$ && $\ell s$, \ $\frac{1}{N_c} \ell t G_c$, \ $\frac{1}{N_c}
\ell^{(2)} gG_c$ \\ \hline

$N_c^{-1}$ && $\frac{1}{N_c} tT_c$, \ $\frac{1}{N_c} \ell S_c$, \
$\frac{1}{N_c} \ell g T_c$, \ $\frac{1}{N_c} S_c^2$, \ $\frac{1}{N_c}
sS_c$, \ $\frac{1}{N_c} \ell^{(2)} sS_c$, \\

&& $\frac{1}{N_c^2} \ell^{(2)} t \{ S_c, G_c \}$, \ $\frac{1}{N_c^2}
\ell^i g^{ja} \{ S_c^j, G_c^{ia} \}$ \\ \hline

$N_c^{-2}$ && $\frac{1}{N_c^2} (\ell S_c)(tT_c)$, \ $\frac{1}{N_c^2} g
S_c T_c$, \ $\frac{1}{N_c^2} \ell^{(2)} S_c S_c$, \ $\frac{1}{N_c^2}
\ell^{(2)} g S_c T_c$, \\

&& $\frac{1}{N_c^2} \{ \ell S_c, s S_c \}$, \ $\frac{1}{N_c^2} (\ell
s) S_c^2$

\end{tabular}
\caption{The 18 linearly independent spin-singlet flavor-singlet
operators for $F=2$, organized by powers of $1/N_c$ in their matrix
elements.  For $F>2$, and ignoring possible coherence in matrix
elements of $T_c^a$, one must include $\frac{1}{N_c^2} t S_c G_c$ and
$\frac{1}{N_c^2} \ell^i g^{ia} S_c^j G_c^{ja}$ in row
$N_c^{-1}$.\label{ops}}
\end{table}
 
\begin{table}

\begin{tabular}{l||cc|cc|cc|cc|c}

& $\langle {\cal O}_1 \rangle$ && $\langle {\cal O}_2 \rangle$ &&
$\langle {\cal O}_3 \rangle$ && $\langle {\cal O}_4 \rangle$ &&
$\langle {\cal O}_5 \rangle$ \\ \cline{2-10}

& $N_c \langle \openone \rangle$ && $\langle \ell s \rangle$ &&
$\frac{1}{N_c} \langle \ell^{(2)} gG_c \rangle$ && $\langle \ell s +
\frac{4}{N_c+1} \ell tG_c \rangle$ && $\frac{1}{N_c} \langle \ell S_c
\rangle$ \\ \hline\hline

$N_{1/2}$ & $N_c$ && $-\frac{1}{3N_c} (2N_c-3)$ && 0 &&
$+\frac{2}{N_c+1}$ && $-\frac{1}{3N_c^2} (N_c+3)$ \\ \hline

$N^\prime_{1/2}$ & $N_c$ && $-\frac 5 6$ && $-\frac{5}{48N_c} (N_c+1)$
&& 0 && $-\frac{5}{3N_c}$ \\ \hline

$N^\prime_{1/2} \,$-$N_{1/2}$ & 0 &&
$-\frac{1}{3}\sqrt{\frac{N_c+3}{2N_c}}$ && $-\frac{5}{48N_c}
\sqrt{\frac{N_c+3}{2N_c}} (2N_c-1)$ && $-\frac{1}{N_c+1}
\sqrt{\frac{N_c+3}{2N_c}}$ &&
$+\frac{1}{3N_c}\sqrt{\frac{N_c+3}{2N_c}}$ \\ \hline

$N_{3/2}$ & $N_c$ && $+\frac{1}{6N_c} (2N_c-3)$ && 0 &&
$-\frac{1}{N_c+1}$ && $+\frac{1}{6N_c^2} (N_c+3)$ \\ \hline

$N^\prime_{3/2}$ & $N_c$ && $-\frac{1}{3}$ && $+\frac{1}{12N_c}
(N_c+1)$ && 0 && $-\frac{2}{3N_c}$ \\ \hline

$N^\prime_{3/2} \,$-$N_{3/2}$ & 0 &&
$-\frac{1}{6}\sqrt{\frac{5(N_c+3)}{N_c}}$ && $+\frac{1}{96N_c}
\sqrt{\frac{5(N_c+3)}{N_c}} (2N_c-1)$ && $-\frac{1}{2(N_c+1)}
\sqrt{\frac{5(N_c+3)}{N_c}}$ &&
$+\frac{1}{6N_c}\sqrt{\frac{5(N_c+3)}{N_c}}$ \\ \hline

$N^\prime_{5/2}$ & $N_c$ && $+\frac 1 2$ && $-\frac{1}{48N_c} (N_c+1)$
&& 0 && $+\frac{1}{N_c}$ \\ \hline

$\Delta_{1/2}$ & $N_c$ && $+\frac 1 3$ && 0 && 0 && $-\frac{4}{3N_c}$
\\ \hline

$\Delta_{3/2}$ & $N_c$ && $-\frac 1 6$ && 0 && 0 && $+\frac{2}{3N_c}$

\end{tabular}

\vskip 2ex
\begin{tabular}{l||@{}cc|cc|cc|c}

& $\langle {\cal O}_6 \rangle$ && $\langle {\cal O}_7 \rangle$ &&
$\langle {\cal O}_8 \rangle$ && $\langle {\cal O}_9 \rangle$ \\
\cline{2-8}

& $\frac{1}{N_c} \langle S_c^2 \rangle$ && $\frac{1}{N_c} \langle sS_c
\rangle$ && $\frac{1}{N_c} \langle \ell^{(2)} sS_c \rangle$ &&
$\frac{N_c+1}{N_c} \langle {\cal O}_4 \rangle + \langle {\cal O}_5
\rangle \! + \!\frac{8}{N_c^2} \langle \ell^i g^{ja} \{ S_c^j,G_c^{ia}
\} \rangle$ \\ \hline\hline

$N_{1/2}$ & $\, \, +\frac{1}{2N_c^2} (N_c+3) \! \!$ &&
$-\frac{1}{4N_c^2} (N_c+3)$ && 0 && $-\frac{1}{3N_c^3}
(17N_c-3)$ \\ \hline

$N^\prime_{1/2}$ & $+\frac{2}{N_c}$ && $+\frac{1}{2N_c}$ &&
$+\frac{5}{6N_c}$ && $+\frac{5}{3N_c^2}$ \\ \hline

$N^\prime_{1/2} \,$-$N_{1/2}$ & 0 && 0 && $+\frac{5}{12N_c}
\sqrt{\frac{N_c+3}{2N_c}}$ && $-\frac{1}{3N_c^2}
\sqrt{\frac{N_c+3}{2N_c}}$ \\ \hline

$N_{3/2}$ & $\, \, +\frac{1}{2N_c^2} (N_c+3) \! \!$ &&
$-\frac{1}{4N_c^2} (N_c+3)$ && 0 && $+\frac{1}{6N_c^3}
(17N_c-3)$ \\ \hline

$N^\prime_{3/2}$ & $+\frac{2}{N_c}$ && $+\frac{1}{2N_c}$ &&
$-\frac{2}{3N_c}$ && $+\frac{2}{3N_c^2}$ \\ \hline

$N^\prime_{3/2} \,$-$N_{3/2}$ & 0 && 0 && $-\frac{1}{24N_c}
\sqrt{\frac{5(N_c+3)}{N_c}}$ && $-\frac{1}{6N_c^2}
\sqrt{\frac{5(N_c+3)}{N_c}}$ \\ \hline

$N^\prime_{5/2}$ & $+\frac{2}{N_c}$ && $+\frac{1}{2N_c}$ &&
$+\frac{1}{6N_c}$ && $-\frac{1}{N_c^2}$ \\ \hline

$\Delta_{1/2}$ & $+\frac{2}{N_c}$ && $-\frac{1}{N_c}$ && 0 &&
$+\frac{4}{3N_c^2}$ \\ \hline

$\Delta_{3/2}$ & $+\frac{2}{N_c}$ && $-\frac{1}{N_c}$ && 0 &&
$-\frac{2}{3N_c^2}$

\end{tabular}

\caption{Matrix elements $\langle {\cal O}_i \rangle_j$ of 9
operators, labeled as ${\cal O}_1, {\cal O}_2, \ldots, {\cal O}_9$,
respectively, that are linearly independent for $N_c=3$.  The third
and sixth rows correspond to off-diagonal matrix
elements.\label{matel1}}
\end{table}
\begin{table}

\begin{tabular}{l||@{}cc|cc|cc|c}

& $\langle {\cal O}_{10} \rangle$ && $\langle {\cal O}_{11} \rangle$
&& $\langle {\cal O}_{12} \rangle$ && $\langle {\cal O}_{13} \rangle$
\\ \cline{2-8}

& $\frac{1}{N_c} \langle \ell g T_c \rangle$ && $\frac{1}{N_c} \langle
t T_c \rangle$ && $\frac{1}{N_c^2} \langle \ell^{(2)} t \{ S_c, G_c \}
\rangle$ && $\frac{1}{N_c^2} \langle (\ell s) S_c^2 \rangle$ \\
\hline \hline

$N_{1/2}$ & $-\frac{1}{12N_c^2} (N_c+3)$ && $-\frac{1}{4N_c^2}
(N_c+3)$ && 0 && $+\frac{1}{6N_c^3} (N_c+3)$ \\ \hline

$N^\prime_{1/2}$ & $+\frac{5}{6N_c}$ && $-\frac{1}{N_c}$ &&
$-\frac{5}{24N_c^2}(N_c+1)$ && $-\frac{5}{3N_c^2}$ \\ \hline

$N^\prime_{1/2} \,$-$N_{1/2}$ &
$+\frac{1}{3N_c}\sqrt{\frac{N_c+3}{2N_c}}$ && 0 && $+\frac{5}{24N_c^2}
\sqrt{\frac{N_c+3}{2N_c}} (N_c+1)$ && $-\frac{2}{3N_c^2}
\sqrt{\frac{N_c+3}{2N_c}}$ \\ \hline

$N_{3/2}$ & $+\frac{1}{24N_c^2} (N_c+3)$ && $-\frac{1}{4N_c^2}
(N_c+3)$ && 0 && $-\frac{1}{12N_c^3} (N_c+3)$ \\ \hline

$N^\prime_{3/2}$ & $+\frac{1}{3N_c}$ && $-\frac{1}{N_c}$ &&
$+\frac{1}{6N_c^2} (N_c+1)$ && $-\frac{2}{3N_c^2}$ \\ \hline

$N^\prime_{3/2} \,$-$N_{3/2}$ &
$+\frac{1}{6N_c}\sqrt{\frac{5(N_c+3)}{N_c}}$ && 0 &&
$-\frac{1}{48N_c^2} \sqrt{\frac{5(N_c+3)}{N_c}} (N_c+1)$ &&
$-\frac{1}{3N_c^2} \sqrt{\frac{5(N_c+3)}{N_c}}$ \\ \hline

$N^\prime_{5/2}$ & $-\frac{1}{2N_c}$ && $-\frac{1}{N_c}$ &&
$-\frac{1}{24N_c^2} (N_c+1)$ && $+\frac{1}{N_c^2}$ \\ \hline

$\Delta_{1/2}$ & $+\frac{1}{6N_c}$ && $+\frac{1}{2N_c}$ && 0 &&
$+\frac{2}{3N_c^2}$ \\ \hline

$\Delta_{3/2}$ & $-\frac{1}{12N_c}$ && $+\frac{1}{2N_c}$ && 0 &&
$-\frac{1}{3N_c^2}$

\end{tabular}

\vskip 2ex
\begin{tabular}{l||cc|cc|cc|cc|c}

& $\langle {\cal O}_{14} \rangle$ && $\langle {\cal O}_{15} \rangle$
&& $\langle {\cal O}_{16} \rangle$ && $\langle {\cal O}_{17} \rangle$
&& $\langle {\cal O}_{18} \rangle$ \\ \cline{2-10}

& $\frac{1}{N_c^2} \langle \{ \ell S_c, s S_c \} \rangle$ && $\,
\frac{1}{N_c^2} \langle (\ell S_c)(tT_c) \rangle$ && $\frac{1}{N_c^2}
\langle g S_c T_c \rangle$ && $\frac{1}{N_c^2} \langle \ell^{(2)} S_c
S_c \rangle$ && $\frac{1}{N_c^2} \langle \ell^{(2)} g S_c T_c \rangle$
\\ \hline\hline

$N_{1/2}$ & $+\frac{2}{3N_c^3} (N_c+3)$ && $+\frac{1}{3N_c^3} (N_c+3)$
&& $+\frac{1}{4N_c^3} (N_c+3)$ && 0 && 0 \\ \hline

$N^\prime_{1/2}$ & $-\frac{5}{3N_c^2}$ && $+\frac{5}{3N_c^2}$ &&
$-\frac{1}{2N_c^2}$ && $+\frac{5}{6N_c^2}$ && $-\frac{5}{6N_c^2}$ \\
\hline

$N^\prime_{1/2} \,$-$N_{1/2}$ &
$-\frac{1}{6N_c^2}\sqrt{\frac{N_c+3}{2N_c}}$ && $\, -\frac{1}{3N_c^2}
\sqrt{\frac{N_c+3}{2N_c}}$ && 0 && $-\frac{5}{6N_c^2}
\sqrt{\frac{N_c+3}{2N_c}}$ && $-\frac{5}{12N_c^2}
\sqrt{\frac{N_c+3}{2N_c}}$ \\ \hline

$N_{3/2}$ & $-\frac{1}{3N_c^3} (N_c+3)$ && $-\frac{1}{6N_c^3} (N_c+3)$
&& $+\frac{1}{4N_c^3} (N_c+3)$ && 0 && 0 \\ \hline

$N^\prime_{3/2}$ & $-\frac{2}{3N_c^2}$ && $+\frac{2}{3N_c^2}$ &&
$-\frac{1}{2N_c^2}$ && $-\frac{2}{3N_c^2}$ && $+\frac{2}{3N_c^2}$ \\
\hline

$N^\prime_{3/2} \,$-$N_{3/2}$ &
$-\frac{1}{12N_c^2}\sqrt{\frac{5(N_c+3)}{N_c}}$ && $\, \,
-\frac{1}{6N_c^2} \sqrt{\frac{5(N_c+3)}{N_c}}$ && 0 &&
$+\frac{1}{12N_c^2} \sqrt{\frac{5(N_c+3)}{N_c}}$ &&
$+\frac{1}{24N_c^2} \sqrt{\frac{5(N_c+3)}{N_c}}$ \\ \hline

$N^\prime_{5/2}$ & $+\frac{1}{N_c^2}$ && $-\frac{1}{N_c^2}$ &&
$-\frac{1}{2N_c^2}$ && $+\frac{1}{6N_c^2}$ && $-\frac{1}{6N_c^2}$ \\
\hline

$\Delta_{1/2}$ & $+\frac{8}{3N_c^2}$ && $-\frac{2}{3N_c^2}$ &&
$-\frac{1}{2N_c^2}$ && 0 && 0 \\ \hline

$\Delta_{3/2}$ & $-\frac{4}{3N_c^2}$ && $+\frac{1}{3N_c^2}$ &&
$-\frac{1}{2N_c^2}$ && 0 && 0

\end{tabular}

\caption{As in Table~\ref{matel1}, for operators labeled as ${\cal
O}_{10}, {\cal O}_{11}, \ldots, {\cal O}_{18}$.\label{matel2}}
\end{table}


\begin{table}
\begin{tabular}{cc|ccc|cccccc|c}
$c_1$ && $c_2$ & $c_3$ && $c_4$ & $c_5$ & $c_6$ & $c_7$ & $c_8$ &&
$c_9$ \\ \hline
$+0.463$ && $-0.036$ & $+0.369$ && $+0.087$ & $+0.086$ & $+0.438$ &
$-0.040$ & $+0.048$ && $+0.001$ \\
$\pm 0.020$ && $\pm 0.041$ & $\pm 0.208$ && $\pm 0.097$ & $\pm 0.080$
& $\pm 0.102$ & $\pm 0.074$ & $\pm 0.172$ && $\pm 0.084$ \\
\end{tabular}
\caption{Operator coefficients in GeV, assuming the complete set of
Table~\protect\ref{matel1}. The vertical divisions separate operators
whose contributions to the baryon masses are of orders $N_c^1$,
$N_c^0$, $N_c^{-1}$, and $N_c^{-2}$, respectively.\label{invert}}
\end{table}


\begin{table}
\begin{tabular}{cccc|ccc}
\multicolumn{7}{c}{Parameters (GeV): $c_1=0.542\pm 0.002$,
$c_2=0.075\pm 0.009$, $c_3=-0.437\pm 0.051 $} \\
\hline\hline
               & Fit       & Exp.\       && & Fit       & Exp.\      
\\ \hline
$\Delta(1700)$ & $1615$ & $1720\pm 50$ &&$N(1520)$ & $1520$ & $1523\pm
8$ \\
$\Delta(1620)$ & $1653$ & $1645\pm 30$ &&$N(1535)$ & $1562$ & $1538\pm
18$\\
$ N(1675) $ & $1677$ & $1678\pm 8$ && $\theta_{N1}$ ({\rm pred}) &
$2.47 \pm 0.04$ & $0.61\pm 0.09$ \\
$ N(1700) $ & $1674$ & $1700\pm 50$ && $\theta_{N3}$ ({\rm pred}) &
$2.65 \pm 0.03$ & $3.04\pm 0.15$ \\
$N(1650)$   &     $1666$ & $1660\pm 20$ &&  & &
\end{tabular}
\caption{Three parameter fit using operators ${\cal O}_{1,2,3}$,
giving $\chi^2/{\rm d.o.f.}=6.89/4=1.72$. The operators included
formally yield the lowest order nontrivial contributions to the masses
in the $1/N_c$ expansion.  Masses are given in MeV, angles in radians.
Experimental data for angles here and below is for comparison purposed
and not used for fitting.}
\label{lowest}
\end{table}


\begin{table}
\begin{tabular}{cccc|ccc}
\multicolumn{7}{c}{Parameters (GeV): $c_1=0.466\pm 0.014$,
$c_2=-0.030\pm 0.039$, $c_3=0.304\pm 0.142$} \\
\multicolumn{7}{c}{$c_4=0.068\pm 0.101$, $c_5=0.062\pm 0.046$,
$c_6=0.424\pm 0.086$}\\  \hline\hline
               & Fit       & Exp.\       && & Fit       & Exp.\      
\\ \hline
$\Delta(1700)$ & $1699$ & $1720\pm 50$ &&$N(1520)$ & $1522$ & $1523\pm
8$ \\
$\Delta(1620)$ & $1643$ & $1645\pm 30$ &&$N(1535)$ & $1538$ & $1538\pm
18$\\
$ N(1675) $ & $1678$ & $1678\pm 8$ && $\theta_{N1}$ ({\rm pred}) &
$0.53 \pm 0.29$ &
$0.61\pm 0.09$ \\
$ N(1700) $ & $1712$ & $1700\pm 50$ && $\theta_{N3}$ ({\rm pred}) &
$3.06 \pm 0.24$ & $3.04\pm 0.15$ \\
$N(1650)$   &     $1660$ & $1660\pm 20$ &&  & &
\end{tabular}
\caption{Six parameter fit using operators ${\cal O}_{1,\cdots,6}$,
giving $\chi^2/{\rm d.o.f.}=0.24/1=0.24$.  Masses are given in MeV,
angles in radians.}
\label{6param}
\end{table}


\begin{table}
\begin{tabular}{cccc|ccc}
\multicolumn{7}{c}{Parameters (GeV): $c_1=0.457\pm 0.005$,
$c_3=0.088\pm 0.198$, $c_6=0.459\pm 0.032$} \\ \hline\hline
               & Fit       & Exp.\       && & Fit       & Exp.\      
\\ \hline
$\Delta(1700)$ & $1678$ & $1720\pm 50$ &&$N(1520)$ & $1525$ & $1523\pm
8$ \\
$\Delta(1620)$ & $1678$ & $1645\pm 30$ &&$N(1535)$ & $1524$ & $1538\pm
18$\\
$ N(1675) $ & $1676$ & $1678\pm 8$ && $\theta_{N1}$ ({\rm pred}) &
$0.11 \pm 0.23$ & $0.61\pm 0.09$ \\
$ N(1700) $ & $1688$ & $1700\pm 50$ && $\theta_{N3}$ ({\rm pred}) &
$3.11 \pm 0.07$ & $3.04\pm 0.15$ \\
$N(1650)$   &     $1668$ & $1660\pm 20$ &&  & &
\end{tabular}
\caption{Three parameter fit using operators ${\cal O}_1$, ${\cal
O}_3$, and ${\cal O}_6$, giving $\chi^2/{\rm d.o.f.}=$
$2.93/4=0.73$.  Masses are given in MeV, angles in radians.}
\label{3param}
\end{table}

\end{document}